# Diffusion time dependence, power-law scaling, and exchange in gray matter

Jonas L. Olesen[1, 2], Leif Østergaard[1], Noam Shemesh[3], Sune N. Jespersen[1, 2,*]

1. Center of Functionally Integrative Neuroscience (CFIN) and MINDLab, Department of Clinical Medicine, Aarhus University, Aarhus, Denmark
2. Department of Physics and Astronomy, Aarhus University, Aarhus, Denmark
3. Champalimaud Research, Champalimaud Centre for the Unknown, Lisbon, Portugal

*Corresponding author. CFIN/MindLab and Dept. of Physics and Astronomy, Aarhus University, Universitetsbyen 3, building 1710, 8000 Aarhus C, Denmark.

E-mail address: sune@cfin.au.dk

## Abstract

Characterizing neural tissue microstructure is a critical goal for future neuroimaging. Diffusion MRI (dMRI) provides contrasts that reflect diffusing spins' interactions with myriad microstructural features of biological systems. However, the specificity of dMRI remains limited due to the ambiguity of its signals vis-à-vis the underlying microstructure. To improve specificity, biophysical models of white matter (WM) typically express dMRI signals according to the Standard Model (SM) and have more recently in gray matter (GM) taken spherical compartments into account (the SANDI model) in attempts to represent cell soma. The validity of the assumptions underlying these models, however, remains largely undetermined, especially in GM. To validate these assumptions experimentally, observing their unique, functional properties, such as the $b^{-1/2}$ power-law associated with one-dimensional diffusion, has emerged as a fruitful strategy. The absence of this signature in GM, in turn, has been explained by neurite water exchange, non-linear morphology, and/or by obscuring soma signal contributions. Here, we present diffusion simulations in realistic neurons demonstrating that curvature and branching does not destroy the stick power-law behavior in impermeable neurites, but also that their signal is drowned by the soma signal under typical experimental conditions. Nevertheless, by studying the GM dMRI signal's behavior as a function of diffusion weighting as well as time, we identify an attainable experimental regime in which the neurite signal dominates. Furthermore, we find that exchange-driven time dependence produces a signal behavior *opposite* to that which would be expected from restricted diffusion, thereby providing a functional signature that disambiguates the two effects. We present data from dMRI experiments in *ex vivo* rat brain at ultrahigh field of 16.4T and observe a time dependence that is consistent with substantial exchange but also with a GM stick power-law. The first finding suggests significant water exchange between neurites and the extracellular space while the second suggests a small sub-population of impermeable neurites. To quantify these observations, we harness the Kärger exchange model and incorporate the corresponding signal time dependence in the SM and SANDI models.

## 1. Introduction

Biophysical modeling of diffusion MRI promises specificity towards microstructural tissue properties, but the crucial step of model validation is challenging due to a paucity of unique signal features and independent methods to estimate model parameters (Novikov et al., 2018, 2019). While the community is slowly reaching consensus concerning modelling water diffusion in white matter (WM), several challenges remain for gray matter (GM) modelling: although several attempts to characterize GM microstructure have been made in the past (Jespersen et al., 2007, 2010; Komlosh et al., 2007; Shemesh and Cohen, 2011; Shemesh et al., 2012; Truong et al., 2014), this area was rightfully coined as a sector of the contemporary *terra incognita* recently (Novikov, 2021). A promising strategy to approach this challenge involves identification of characteristic functional dependencies that act as model signatures and allow discrimination between competing models (Fieremans et al., 2016; McKinnon et al., 2017; Veraart et al., 2019; Lee et al., 2020b, 2020a, 2021). An example is the $b^{-1/2}$ power-law of the powder averaged signal from zero radius cylinders ("sticks") at large b-values (Veraart et al., 2019). This behavior was recently observed in brain WM (McKinnon et al., 2017; Veraart et al., 2019, 2020), providing strong support for a stick compartment, presumably axons, of highly anisotropic spaces with negligible radial diffusivity (Behrens et al., 2003; Kroenke et al., 2004; Jespersen et al., 2007). This has been an important validation step for the "Standard Model" (SM) of diffusion, an umbrella term for a class of models that consist of an axonal stick compartment and an extra-axonal water compartment (Novikov et al., 2019), considered now widely to be an adequate description of WM diffusion data at sufficiently long diffusion times.

In contrast, the stick power-law has not been observed in GM (McKinnon et al., 2017; Veraart et al., 2020). This was first ascribed to non-negligible water exchange across the neurite membrane (McKinnon et al., 2017; Veraart et al., 2020). Later, it was suggested to break down due to neurite curvature (Özarslan et al., 2018). Finally, a separate signal component may be needed to account for cell bodies (soma) (Palombo et al., 2020). Such a component would tend to obscure the neurites' behavior and explain the power-laws absence in GM. This hypothesis is supported experimentally by the agreement between GM data and the soma and neurite density imaging (SANDI) model, which extends the SM with a soma compartment (Palombo et al., 2020), and numerically by simulations in microscopy reconstructed neurons (Fang et al., 2020; Olesen and Jespersen, 2020).

Here we set out to discriminate these explanations for the GM power-law's absence. Using simulations of diffusion in realistic neurons, we validate that impermeable neurites do exhibit the stick power-law even with realistic morphology. While somas will typically obscure this, gradient pulse times can be carefully chosen to make the neurite signal dominate at large diffusion weighting revealing its potential power-law behavior. Analyzing the time dependence of exchange and non-exchange diffusion models, we identify a separate functional behavior which provides a signature for dominant exchange, namely whether the signal increases or decreases as a function of diffusion time at fixed diffusion weighting. Data from 4 *ex vivo* rat brains is presented with up to very large diffusion weighting ($b_{max}$ = 100 ms/µm²) and pulse times chosen in accordance with the simulations. The observed time dependence indicates substantial water exchange and is accounted for by incorporating exchange in the SM. We also observe an apparent stick power-law in GM consistent with a small sub-population of impermeable neurites, plausibly myelinated axons. Furthermore, the data is consistent with a distinct soma contribution suggesting that both exchange and somas contribute to the typical lack of a GM power-law.

## 2. Methods

### 2.1. Models

We interpret the powder-averaged GM signal as having three main sources: extracellular water, neurite water residing in dendrites and axons, and soma water. Other sources may contribute but are assumed to have sufficiently similar diffusion properties to one of the main sources that they are implicitly accounted for. For example, aquaporins (water channel proteins) in glial cells (Manley et al., 2004) possibly mediates sufficiently fast water exchange with the extracellular space to make the two indistinguishable from a single effective signal component. The models generally include an offset $f_{im}$ to account for a possible contribution from immobile water as reported *ex vivo* (Kroenke et al., 2006; Alexander et al., 2010; Williamson et al., 2019). We compare two classes of models: one excluding and one including water exchange.

*2.1.1. SANDI*

Consider first the case of negligible exchange. The recently proposed GM model SANDI (Palombo et al., 2020) describes each signal source in the Gaussian phase approximation (GPA) and attribute isotropic diffusivity to soma and extracellular water, while the neurites are modelled as narrow cylinders (sticks) with effectively zero radial diffusivity (see Fig. 1). As an extension, the extracellular diffusion can be straightforwardly relaxed to be anisotropic (Jespersen et al., 2010). However, we found that the effect on fitting quality and parameter estimates was small, and therefore limit the analysis to an isotropic compartment (Jespersen et al., 2007). Therefore, the powder-averaged signal is given by (Palombo et al., 2020)

$$\overline{S}(b,\Delta,\delta)/S_0 = f_e \exp(-bD_e) + f_n \sqrt{\frac{\pi}{4bD_n}} \mathrm{erf}(\sqrt{bD_n}) + f_s \exp(-bD_s(\Delta,\delta)) + f_{im}, \quad (1)$$

where subscripts $e$, $n$, $s$, and $im$ refer to extracellular, neurites, somas, and immobile respectively. Note, the weights $f_i$ are defined differently than in (Palombo et al., 2020): each $f_i$ is a fraction of the total signal and therefore referred to as signal fractions throughout. The diffusivity $D_s$ is related to the soma size through the GPD approximation (Balinov et al., 1993; Stepišnik, 1993). It also depends on the free diffusivity $D_0$ but typically only weakly (Li et al., 2017). In our case, varying it between 1.5 and 3 µm²/ms affect fitting quality very little and has only a minor effect on parameter estimates. We therefore fix it to 2 µm²/ms.

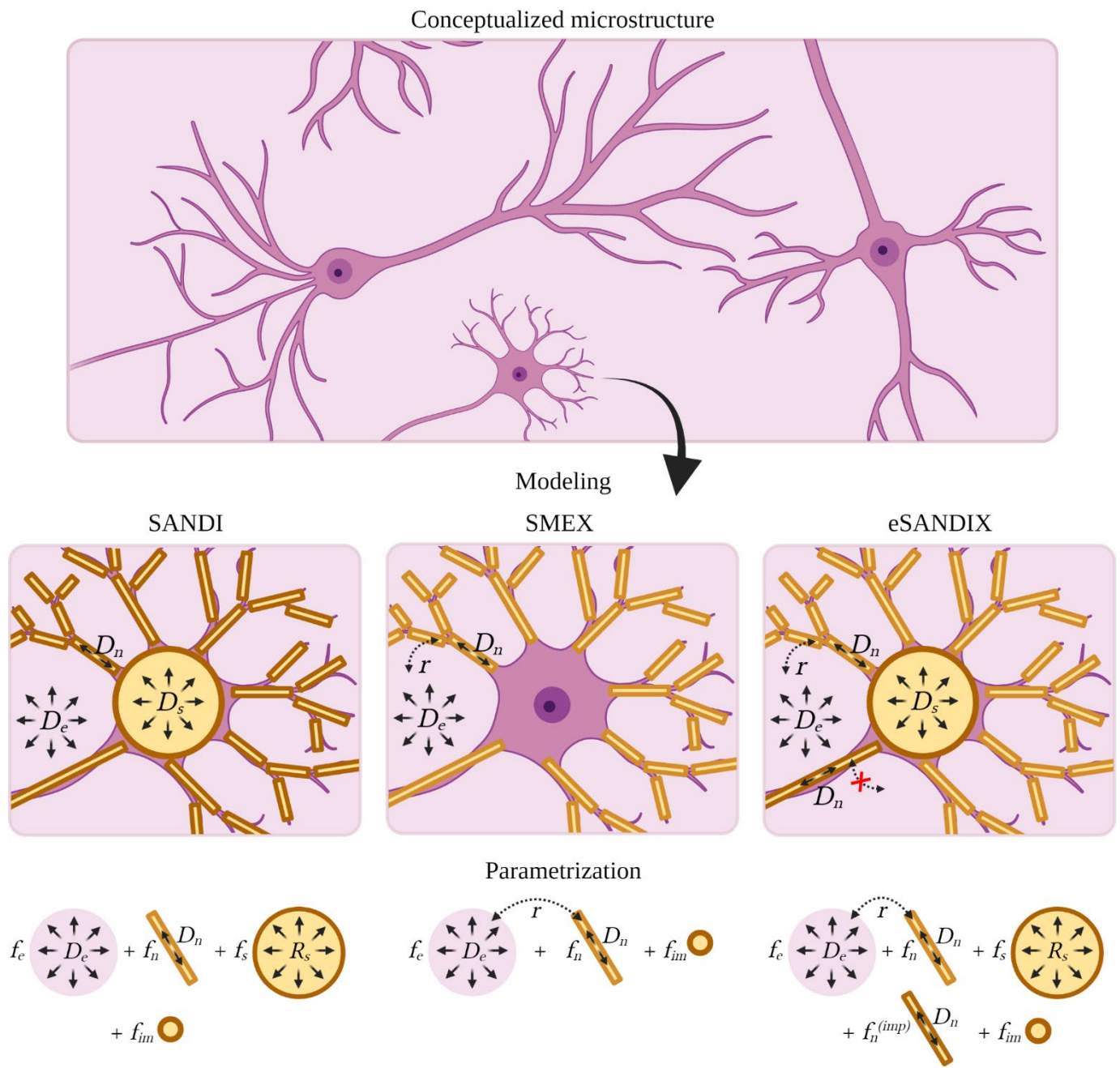

Fig. 1. Graphical representation of the considered models. The SANDI model represents the dMRI signal with sticks, spheres and an isotropic Gaussian component. These can tentatively be assigned to neurites, somas, and extracellular water. The SMEX model excludes spheres but includes exchange between the sticks and the isotropic Gaussian component. Two extensions to SMEX are considered: including spheres/somas and introducing a sub-population of sticks which do not exchange.

Regarding the regime of validity, (Palombo et al., 2020) found exchange between somas and neurites to be insignificant at diffusion times below 20 ms, and that neurites can be modelled as sticks despite branching and finite processes (Palombo et al., 2020). Our simulations support this and include undulations (see section 3.1). We argue that the GPA is applicable for somas even with very large b-values because the soma signal is already substantially attenuated once it deviates from the GPA at which point the overall signal is dominated by neurites. This was verified by comparing with the exact sphere signal from the multiple correlation function (MCF) approach (Callaghan, 1997; Grebenkov, 2008).

*2.1.2. SMEX, SANDIX, and eSANDIX models*

Three exchange models are considered: SMEX, SANDIX, and eSANDIX in order of complexity. With SMEX we refer to the SM with exchange between neurites and extracellular water. It was previously invoked to explain the lack of a stick power-law in GM (Jelescu and Novikov, 2020; Veraart et al., 2020) and in a concurrent study as Neurite Exchange Imaging (NEXI) (Jelescu et al., 2021). Somas are not explicitly included but presumably implicitly accounted for by the extracellular compartment; the model thus relies on similar diffusion properties of those two components. A natural extension is therefore to add a soma compartment resulting in SANDI with exchange (SANDIX). The somas and neurites are assumed to have negligible exchange on the same grounds as in SANDI (Palombo et al., 2020). Exchange between somas and extracellular water is also neglected because of the negligible surface to volume ratio of somas relative to neurites. Finally, introducing a sub-population of impermeable neurites (presumably myelinated axons) enables the model to reproduce the data's apparent stick power-law (see section 2.2). We refer to this model as eSANDIX.

Exchange is implemented using the Kärger model (Kärger, 1985) generalized to arbitrary gradient profiles (Ning et al., 2018). It omits the spatiotemporal coupling of the real diffusion-exchange process by describing it in terms of two Gaussian populations coupled by an effective rate parameter. This is a good approximation when exchange is barrier limited (Fieremans et al., 2010) – i.e. the spins fully explore each compartment before escaping. The signal is calculated by numerically solving the generalized rate equations

$$\frac{\partial}{\partial t}\begin{bmatrix} S_1(t) \\ S_2(t) \end{bmatrix} = \left( \begin{bmatrix} -r_1 & r_2 \\ r_1 & -r_2 \end{bmatrix} - q^2(t) \begin{bmatrix} D_1 & 0 \\ 0 & D_2 \end{bmatrix} \right) \begin{bmatrix} S_1(t) \\ S_2(t) \end{bmatrix}, \qquad (2)$$

where $D_i$ is the diffusivity and $r_i$ is the exchange rate of compartment $i$. This is valid for any given gradient profile $g(t)$, with $q(t) \equiv \gamma \int_0^t g(t')\,dt'$. The exchange rates are dependent parameters, as they satisfy detailed balance, $r_1 f_1 = r_2 f_2$. The observed signal is the sum of $S_1$ and $S_2$ evaluated at the end of the encoding.

Consider a system consisting of a single stick (neurite) of orientation $\hat{n}$ exchanging with the extracellular water. For gradient direction $\hat{g}$, the apparent neurite diffusivity is $D_n \varepsilon^2$, where $\varepsilon \equiv \hat{g} \cdot \hat{n}$. The signal for that particular gradient direction is thus calculated by integrating

$$\frac{\partial}{\partial t}\begin{bmatrix} S_1(t) \\ S_2(t) \end{bmatrix} = \left( \begin{bmatrix} -r_n & r_e \\ r_n & -r_e \end{bmatrix} - q^2(t) \begin{bmatrix} D_n \varepsilon^2 & 0 \\ 0 & D_e \end{bmatrix} \right) \begin{bmatrix} S_1(t) \\ S_2(t) \end{bmatrix}. \tag{3}$$

The powder-averaged signal can be estimated using Gauss-Legendre quadrature with respect to $\varepsilon$ since this expression is axisymmetric. The actual system of interest generally contains many neurites with differing orientations, which in principle requires the rate equation to be extended with a component for each. For tractability, we assume that each neurite only exchanges with its local extracellular water in the sense that each spin visits only one neurite (or multiple parallel neurites) over the course of the encoding. This is a first approximation and effectively partitions the system into "cells" with intra-compartmental exchange but which do not exchange with each other – conceptually similar to the partitioning into non-exchanging fascicles in the SM. We note that the validity of this approximation could be strained with fast exchange. For instance, with the reported eSANDIX parameter estimates (Table 1), a spin's probability for exchanging from one neurite to the extracellular space and reenter a neurite (potentially a differently oriented one) is roughly 60% per 10 ms. In terms of signal, the cells differ only in orientation so their powder-averages are identical. Therefore, the total combined neurite and extracellular signal up to a volume factor is calculated by solving Eq. (3) and powder-averaging.

The total SMEX signal is obtained by adding $f_{im}$ (see Fig. 1). For the SANDIX and eSANDIX signals, a sphere and for eSANDIX also an impermeable stick compartment is added. The diffusivities of the permeable and impermeable sticks are constrained to be equal.

## 2.2. Power-law-scaling for model discrimination

Consider the signal for SANDI in Eq. (1). In the limit of large b-values, the extracellular and soma signals are negligible relative to the neurite signal due to their exponential attenuation – in practice, the exponential behavior will only be accurate within a b-value range but beyond that, the neurite signal will likely dominate rendering the other compartments' specifics irrelevant. The neurite signal converges to a $b^{-1/2}$ power-law as the error function approaches 1 exponentially. SANDI thus explicitly includes this signal feature.

The power-law is broken if neurite exchange is appreciable (Veraart et al., 2019), as is intuitive since it enables deviation from one-dimensional diffusion. This is explicitly seen in the narrow pulse limit where the Kärger model can be solved analytically. The solution has been presented prior to this by (Kärger, 1985; Fieremans et al., 2010) and many others

$$S = f'_1 e^{-a_1} + f'_2 e^{-a_2}, \tag{4}$$

$$a_{1,2} \equiv \frac{1}{2}\left(bD_1 + bD_2 + tr_1 + tr_2 \pm \sqrt{(bD_1 - bD_2 + tr_1 - tr_2)^2 + 4tr_1 tr_2}\right), \tag{5}$$

$$f'_{1,2} \equiv \pm \frac{b\overline{D} - a_{2,1}}{a_1 - a_2}, \tag{6}$$

where $\overline{D} \equiv f_1 D_1 + f_2 D_2$. The signal $S(\varepsilon)$ for a specific orientation is obtained by substituting $D_1$ for the neurite diffusivity $D_n \varepsilon^2$ in said direction. The powder-averaged signal is obtained by integrating over $\varepsilon$, which is well-approximated by a power-series in $b^{-1/2}$ in the limit of large b-values (Veraart et al., 2020). The series can be obtained by using Laplace's method and Taylor expanding in the limit $b \to \infty$. To order $b^{-3/2}$, we find

$$\overline{S} \equiv \frac{1}{2}\int_{-1}^{1} S(\varepsilon)\, d\varepsilon \sim f_n \sqrt{\frac{\pi}{4bD_n}} e^{-tr_n}\left(1 + tr_n \frac{2 + tr_e}{bD_e} + \cdots\right). \tag{7}$$

The rate of convergence for the expression depends on the exchange rates with slower exchange leading to faster convergence and vice versa. The regular power-law is reproduced by nulling the exchange. Notably, exchange has two effects: exponential attenuation in proportion to the exchanged volume of neurite water $tr_n$, and corrections to the $b^{-1/2}$ power-law of higher-order in $b$. The expression still approaches a $b^{-1/2}$ power-law at sufficiently large b-values, but it happens slower than exponentially. Consequently, SMEX cannot reproduce the power-law in practice if exchange is appreciable. This can be mitigated by introducing a non-exchanging stick compartment, which amplifies the weight of the $b^{-1/2}$ term relative to higher-order terms.

### 2.3. Diffusion time dependence as an exchange signature

The diffusivity generally equals the free diffusivity at very short times and decreases as a function of time towards a microstructure-dependent effective diffusivity in the long time limit due to hindrances or restriction (Novikov and Kiselev, 2010; Novikov et al., 2014). In SANDI, we neglect compartmental kurtosis and higher-order effects (Henriques et al., 2020). The compartmental signal in a specific direction is therefore on the form $f_i \exp(-bD_i(t))$. At fixed b-value, this is an increasing function of time because the diffusivity decreases with time:

$$\frac{d}{dt} f_i e^{-bD_i(t)} = -b f_i e^{-bD_i(t)} \frac{dD_i(t)}{dt} \geq 0 . \tag{8}$$

Consequently, the powder-averaged signal of SANDI or any non-exchanging Gaussian multicompartment model also increases with time at fixed b-value because it is a sum of terms obeying the inequality.

In the Kärger model, the diffusivities are time independent, but exchange produces time dependence in the signal. Even so, the overall diffusivity is constant (Fieremans et al., 2010). Accordingly, the signal's time dependence increases from none at zero b-value to being substantial at large b-values. Crucially, the model exhibits the opposite time dependence of Eq. (8). This is evident from Eq. (7) for large b-values, but the Kärger signal is in fact generally (for any $t$, $b$, and choice of model parameters) a decreasing function of time at fixed b-value:

$$\frac{d}{dt}\bar{S} = -f_1 r_1 (bD_2 - bD_1)^2 \exp\left(-\frac{a_1 + a_2}{2}\right) \frac{s \cosh(s) - \sinh(s)}{s^3} \leq 0 , \tag{9}$$

$$s \equiv \sqrt{(bD_1 - bD_2 + tr_1 - tr_2)^2 + 4tr_1 tr_2} . \tag{10}$$

Exchange and structure driven time dependence thus have opposite signs, producing a signature for deciding the dominant effect. Both effects will generally contribute; in SANDIX, for instance, the neurites exchange with extracellular water while the somas obey Eq. (8). In practice, dominant exchange will cause a signal decrease with increasing time, but that does not rule out a smaller contribution from structure and vice versa.

2.4. Simulations

We employ Monte Carlo diffusion simulations in microscopy reconstructed human pyramidal cells to investigate the signal behavior of GM neurons with realistic geometries but impermeable membranes. In particular, the simulations serve to support that the neurite signal is well-described by sticks providing the basis for the considered models.

Ten reconstructions were obtained from NeuroMorpho.org (Ascoli et al., 2007) under the Allen Brain Atlas (Koch and Jones, 2016) in swc format (connected nodes with associated radii). The swc files were converted to three-dimensional tetrahedron meshes using the Blender addons from the Mcell team (Stiles et al., 1996; Stiles and Bartol, 2000; Kerr et al., 2008). Afterwards, diffusion simulations were carried out by initializing $10^6$ particles uniformly distributed in the geometry. The particle positions were updated each time step using three-dimensional Gaussian steps characterized by a typical step size σ of 0.01 µm, substantially smaller than any cell dimensions. The length of the time steps was set accordingly to reproduce a free diffusivity of 2 µm²/ms. The restricting geometry was enforced by detecting whenever a particle would pass through a tetrahedron face. If no neighboring tetrahedron shared that face, the face was part of the neuron membrane and the collision was handled as an elastic collision with a plane (reflecting boundaries). Otherwise, the particle was allowed to pass to the neighboring tetrahedron.

It is useful to consider the neurite and soma signal contributions separately. These are not strictly separable at finite diffusion time because particles in the soma can diffuse into the neurites and vice versa. Recognizing this, we define the neurite signal as stemming from particles starting at a distance of more than $R + 2\sqrt{2D_0 t}$ from the soma, and the remainder of the particles make of the soma signal. Particles contributing to the neurite signal thus have very low probability of encountering the soma, while a fraction of those contributing to the soma signal will only partially or not at all encounter the soma. The neurite signal, which is of main interest, is thus very representative of neurites while the soma signal also has a small neurite contribution. However, the "in between" particles only make up roughly 5% of the soma particles.

## 2.5. Data

All animal experiments were preapproved by the competent institutional and national authorities and strictly adhered to European Directive 2010/63.

Specimens were transcardially perfused, immersed in 4% Paraformaldehyde (PFA) solution (24 h), and washed in Phosphate-Buffered Saline (PBS) solution (48 h). A hemisphere was isolated and placed in a 5 mm NMR tube with Fluorinert kept at 37 °C and scanned using a 16.4 T Bruker Aeon scanner with a Micro5 probe (producing gradients up to 3000 mT/m). After minimum 1 h of temperature equilibration and imaging adjustments, diffusion data was recorded with EPI readout (three shots, 268 kHz bandwidth), Field of View 1.1x1.1 cm$^2$, matrix size 80x80 (13 slices with 1000 µm thickness each), in plane resolution 138x138 µm$^2$, TR = 4 s, TE = 30 ms, and thirty gradient directions uniformly distributed on a hemisphere. Gradient pulse width δ = 4.5 ms and separation Δ = 16 ms were chosen to facilitate the observation of a possible stick power-law in GM (see section 3.1). Therefore, large b-values were densely sampled for this δ, Δ combination. To investigate the dependence on diffusion time, data was also recorded at Δ = 11 and 7.5 ms. The b-values were b = 0.1, 0.5, 1, 2, 3, 4, and 5 ms/µm$^2$ and with larger b-values chosen linearly in b$^{-1/2}$: In units of µm/ms$^{1/2}$, from 0.4 to 0.15 for Δ = 7.5 ms, from 0.4 to 0.125 for Δ = 11 ms, and from 0.4 to 0.1 for Δ = 16 ms. The spacing was 0.025 µm/ms$^{1/2}$, resulting in respectively eleven and twelve such b-values for Δ = 7.5 ms and 11 ms. For Δ = 16 ms, the spacing was halved between 0.3 and 0.1 µm/ms$^{1/2}$, resulting in a total of 21 such b-values. The number of averages was ten for b ≤ 25 ms/µm$^2$ and thirty otherwise. One image without diffusion weighting was acquired for each b-shell for drift correction. The supporting datasets slightly differ in choices of Δ and b-values as detailed in appendix A.

The data was denoised using MP-PCA (Veraart et al., 2016) and corrected for Rician bias using the inversion technique of (Koay and Basser, 2006) with the noise variance estimated from the mean signal in a tissue-free region. We performed rigid co-registration based on cross-correlation (Guizar-Sicairos et al., 2008) and corrected for signal magnitude drift using the non-diffusion-weighted images (Vos et al., 2017).

Our analysis focuses on a single ROI chosen to be representative of the GM signal. We also include an ROI representative of WM and one GM ROI with minimal myelin for comparison. The ROIs are depicted in Fig. 2 and are placed in the cortex, corpus callosum, and amygdala. The cortical ROI is subdivided into N=6 sub-ROIs to assess parameter variation. In the ensuing analysis, we refer to the voxel-averaged signals from the cortex and corpus callosum as the GM and WM signals.

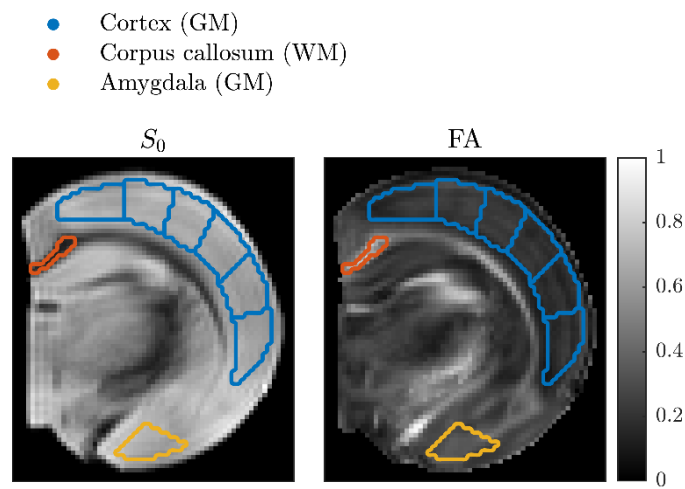

Fig. 2. GM and WM ROIs superimposed on maps of signal magnitude at b = 0 (left) and fractional anisotropy (right) estimated from a DKI fit to the subset of data with b ≤ 3 ms/μm$^2$. The ROIs in cortex and corpus callosum are referred to as the GM and WM ROIs.

2.6. Model parameter estimation

Model parameters were estimated using non-linear least-squares fitting. The subdivision of the cortical ROI (Fig. 2) was leveraged to asses parameter variation and fitting stability: parameters were for each sub-ROI using 1000 random initializations to find each global minimum and other minima with comparable fitting quality. The initial parameters were uniformly distributed within large but reasonable intervals; for example between 0 and 3 $\mu m^2$/ms for the diffusivities.

For increased efficiency, variable projection (Golub and Pereyra, 2003; Farooq et al., 2016; Fadnavis et al., 2019) was used to eliminate the signal fractions from the non-linear fitting, thereby reducing the dimensionality. The technique exploits the linearity in the signal fractions: throughout the iterative fitting procedure, after each update of the non-linear parameters, the optimal signal fractions can be efficiently estimated with linear least-squares fitting.

## 3. Results

### 3.1. Simulations

Fig. 3 shows the simulated signal as well as the separate soma and neurite signal contributions for pulse times matching the experiment (panel a) and those of (Veraart et al., 2020) (panel b) for which a GM stick power-law was not observed. The simulated signals demonstrate that impermeable neurites behave approximately as sticks at feasible experimental settings despite realistic curvature and branching. These deviations from ideal sticks result in minor transverse mobility and consequently an apparent power-law exponent slightly above 1/2. The neurite contribution is generally obscured by somas, but, at very large b-values, the soma contribution is sufficiently attenuated for the neurites to dominate revealing the underlying power-law. This stick signature is thus observable at very large diffusion weighting, which is conceptually similar to the need for attenuating the extra-axonal signal in order to observe the power-law in WM (McKinnon et al., 2017; Veraart et al., 2019, 2020), but potentially requires larger b-values depending on soma sizes and the ensuing attenuation. As exemplified by comparing panels a) and b), the gradient pulse times can be chosen to reduce the minimum required diffusion weighting.

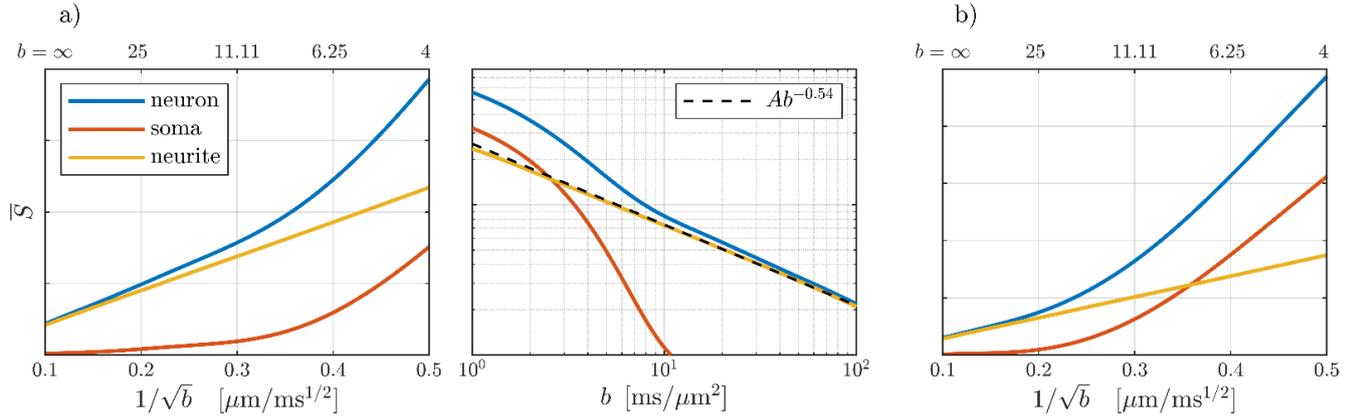

Fig. 3. The simulated signal as well as the separate soma and dendrite contributions for two combinations of pulse times: a) δ = 4.5 ms and Δ = 16 ms matching the experiment with the signal shown as a function of $b^{-1/2}$ (left) and in a double-log plot (middle) where the power-law shows as a straight line. b) δ = 13 ms and Δ = 30 ms matching those of (Veraart et al., 2020).

We define the apparent power-law exponent as the absolute value of the slope obtained from linearly fitting to the log-log signal curve from b = 10 to 100 ms/μm². This exponent is shown as a function of pulse times in Fig. 4 for the total signal and for the neurite contribution. The neurites follow the stick power-law well for a broad range of gradient pulse times. This is least fulfilled at short Δ and longer δ, but generally the neurite exponent approaches -1/2 for long Δ. Contrastingly, the soma contribution increases with Δ and δ as seen in Fig. 4 by the behavior of the exponent of the total signal. To best facilitate the observation of the neurite power-law in GM, δ should be small, while Δ has an optimum between 10-20 ms dependent on δ.

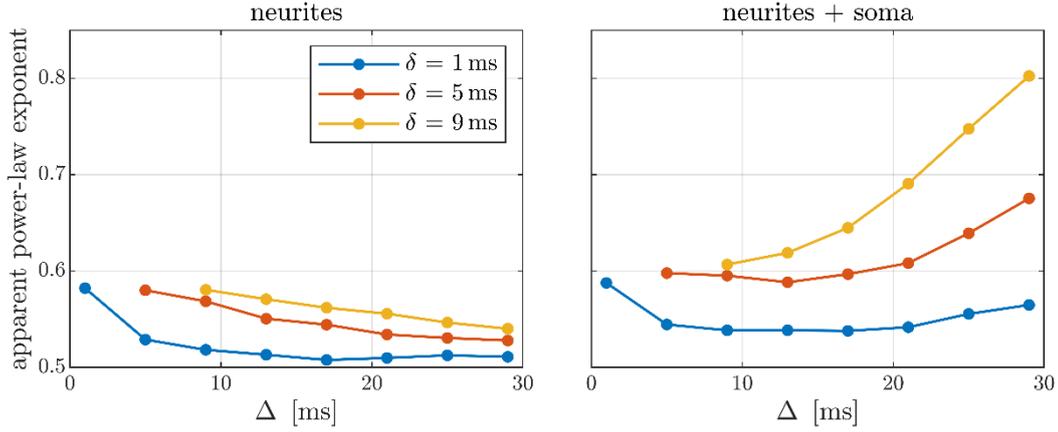

Fig. 4. The apparent power-law exponent for the neurite signal contribution (left) and the total neuron signal (right) as a function of pulse separation Δ for different pulse durations δ. The apparent exponent is here defined as the absolute value of the slope obtained from fitting to the log-log signal curve from b = 10 to 100 ms/µm².

In practice, the achievable Δ and δ depend on the available gradient strength, constrained by $b_{max} = g_{max}^2 \delta^2 (\Delta - \delta/3)$ (ignoring ramp times) for a specific b-value target. Consequently, short Δ entails long δ and vice versa, resulting in a trade-off between minimizing the neurites' deviation from sticks and the soma contribution. Fig. 5 illustrates this for the employed scanner: neurite exponent is minimized at short δ, long Δ but the soma contribution is minimized at longer δ, shorter Δ. The resulting optimum is approximately δ/Δ = 4/20 ms. In the right panel of Fig. 5, the achievable apparent exponent is shown as a function of the available gradient strength: the exponent increases rapidly with decreasing gradient strength rendering the power-law unobservable with typical gradients.

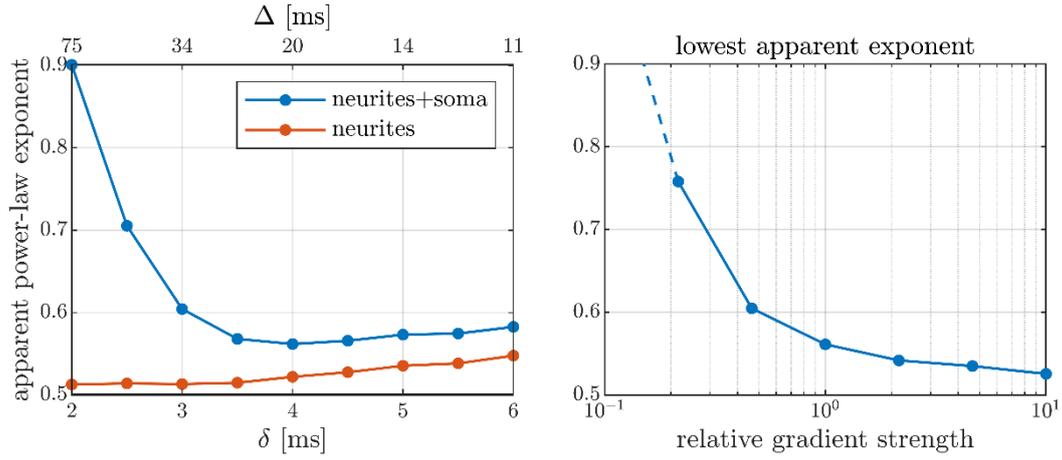

Fig. 5. Left panel: Apparent power-law exponent for the neurite contribution and the total signal as functions of δ and Δ constrained by a target b = 100 ms/µm² with the employed scanner. Right panel: The lowest apparent exponent achievable (optimizing pulse times) as a function of available gradient strength relative to the employed scanner ($g_{max}$ = 3000 mT/m).

### 3.2. Power-law scaling and time dependence in *ex vivo* brain

Experiments were performed on N = 4 rats. Here, data from a single rat is presented. The results from the remaining datasets are fully consistent and presented in appendix A.

The GM mean diffusivity and kurtosis as functions of diffusion time are shown in Fig. 6. The diffusivity decreases but the effect is subtle (≈ 5% from 7.5 to 16 ms) relative to the kurtosis' decrease (≈ 32% from 7.5 to 16 ms).

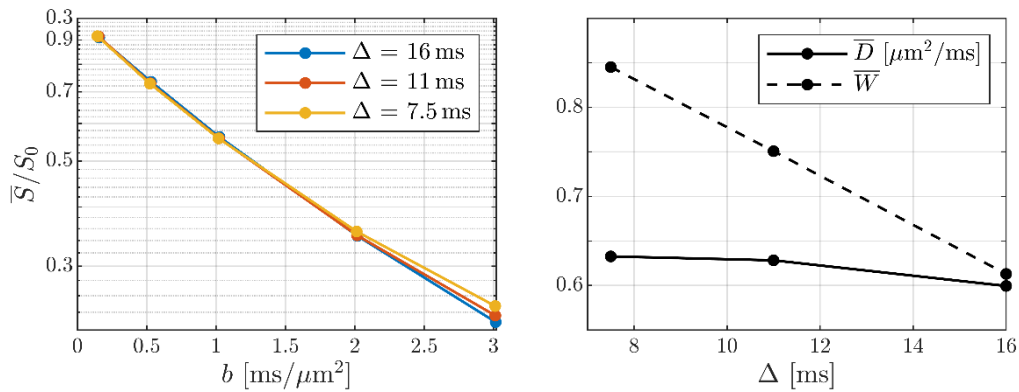

Fig. 6. Left: the GM signal at small b. Right: the corresponding mean diffusivity and kurtosis. Parameters were estimated for each diffusion time with DKI applied to the subset of data with b ≤ 3 ms/µm².

The GM and WM signals are shown in Fig. 7 as functions of $b^{-1/2}$: the WM signal clearly reproduces the previously reported power-law and shows no appreciable time dependence. The GM signal behaves consistently with a $b^{-1/2}$ power-law for b > 25 ms/µm² at pulse separation Δ = 16 ms, and shows substantial time dependence with the signal being decreasing with time at fixed b-value.

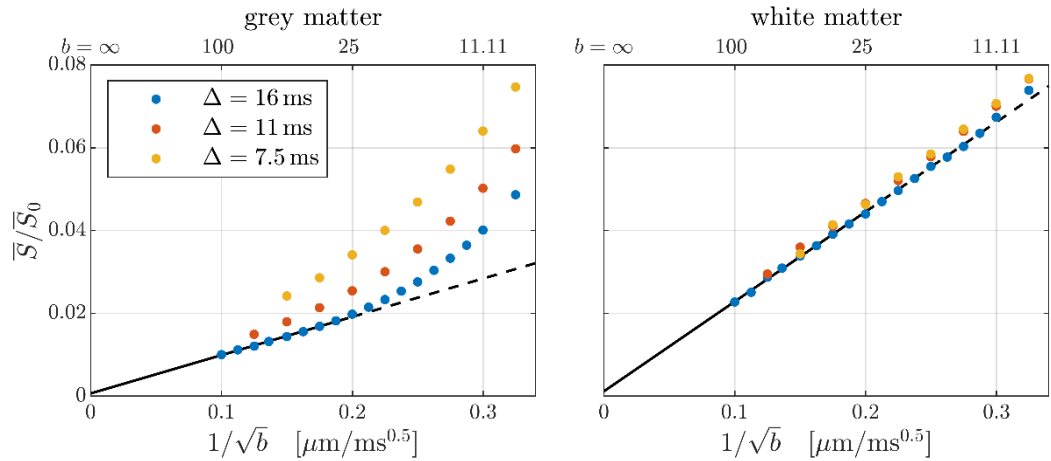

Fig. 7. The GM and WM signals at large b-values (ROIs given in Fig. 2). The black lines show linear fits to the subset of data with b ≥ 25 ms/µm² and Δ = 16 ms – lines are solid for b ≥ 25 ms/µm² and otherwise dashed.

The time dependence of the GM signal is consistent with SMEX as illustrated in Fig. 8: model parameter estimates for Δ = 16 ms is used to predict the signal at Δ = 7.5 and 11 ms. The opposite time dependencies of SMEX and SANDI are clearly illustrated. The data also emphasizes that the dMRI signal at a single diffusion time can make fitting quality alone a problematic metric for model discrimination, as the data with Δ = 16 ms is overall well described by both SANDI and SMEX (Novikov et al., 2018).

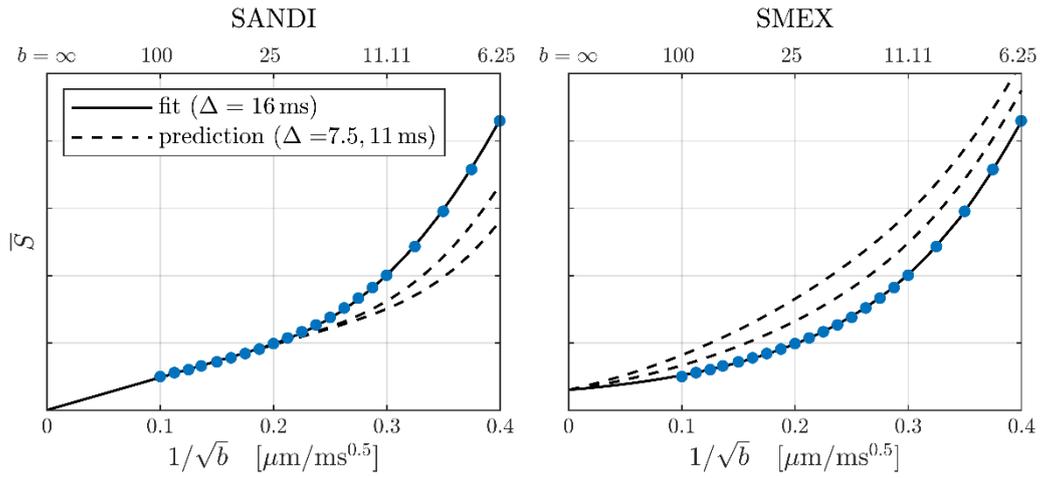

Fig. 8. Both panels show the GM signal for Δ = 16 ms (ROI given in Fig. 2). Fits of SANDI (left) and SMEX (right) are shown with solid curves, while dashed curves show the predicted signal at Δ = 11 and 7.5 ms (using parameters obtained from the fitting). Parameter estimates are provided for completeness but note that the fits employed only the subset of data with Δ = 16 ms. For SANDI: $f_e$ = 64%, $D_e$ = 0.7 μm²/ms, $f_n$ = 17%, $D_n$ = 2.5 μm²/ms, $f_s$ = 19%, $R_s$ = 5.4 μm, and $f_{im}$ = 0. For SMEX: $f_e$ = 54%, $D_e$ = 0.6 μm²/ms, $f_n$ = 45%, $D_n$ = 1.9 μm²/ms, $\tau_n$ = 8.1 ms, and $f_{im}$ = 0.6%.

Fig. 9 shows fits to the full dataset (i.e., the three diffusion times together) of SMEX and eSANDIX. SANDI is not fitted since it is incompatible with the data's time dependence. Contrarily, SMEX describes the time dependence well but does not fully reproduce the apparent power-law for Δ = 16 ms, as evident from the divergence of the fit from the data at the highest b-values (lowest $b^{-1/2}$). Parameter estimates are given in Table 1: SMEX suggests similar intra- and extra-neurite diffusivities between 1.0-1.5 μm²/ms with the neurites constituting two thirds of the signal. The exchange is fast with mean neurite residence time estimated to $\tau_n$ ≈ 4 ms.

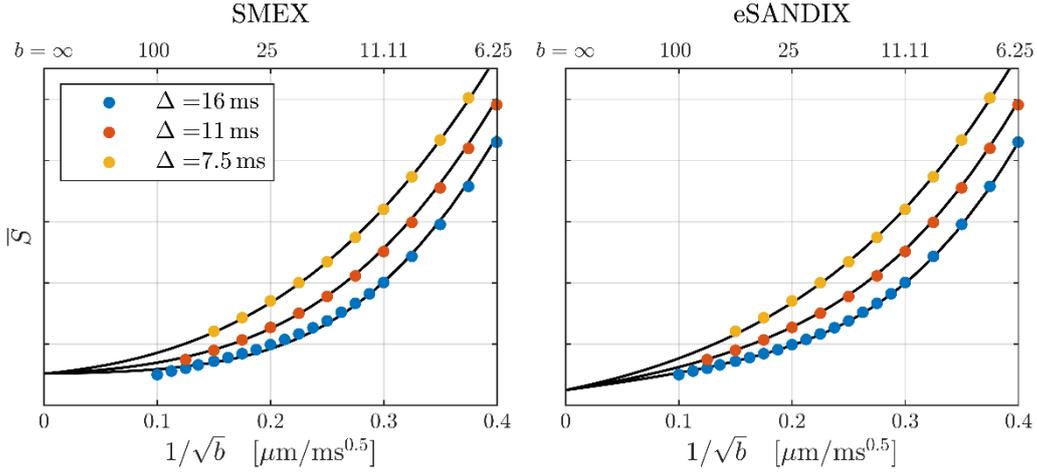

Fig. 9. Both panels show the GM signal (ROI given in Fig. 2). The curves show fits of SMEX (left) and eSANDIX (right). Parameter estimates are given in Table 1.

|  | $f_e$ | $D_e$ | $f_n$ | $f_n^{(imp)}$ | $D_n$ | $\tau_n$ | $f_s$ | $R_s$ | $f_{im}$ |
|---|---|---|---|---|---|---|---|---|---|
| SMEX | 30-34 | 1.0-1.1 | 65-69 | - | 1.1-1.5 | 3.8-4.5 | - | - | 0.8-1.2 |
| SANDIX | 29-37 | 0.9-1.0 | 47-59 | - | 0.6-1.0 | 3.8-5.1 | 10-18 | 11.8-13.5 | 0.9-1.1 |
| eSANDIX | 37-45 | 0.8-0.9 | 36-44 | 1-5 | 0.4-0.6 | 3.6-4.4 | 13-20 | 12.4-13.5 | 0.4-0.7 |
| SANDIX* | 12-17 | 1.9-2.7 | 73-78 | - | 0.7-1.1 | 3.2-4.0 | 7-10 | 6.5-8.2 | 0.9-1.2 |
| eSANDIX* | 13-22 | 1.9-3.0 | 66-70 | 2-6 | 0.5-0.8 | 3.0-3.6 | 8-13 | 6.3-8.1 | 0.5-0.7 |

Table 1. Model parameter estimates given as min-max ranges from fitting to each sub-ROI in the cortex (see Fig. 2). The asterisks * mark alternative minima with comparable fitting quality. Diffusivities are given in μm²/ms, signal fractions in %, $\tau_n$ in ms, and $R_s$ in μm.

Extending from SMEX to SANDIX (fit not shown) increases the fitting quality significantly based on Bayesian information criterion (ΔBIC > 10) according to typical model selection recommendations (Kass and Raftery, 1995), but does not qualitatively improve the fit in terms of time dependence or large b-value behavior. Model parameter estimates are given in Table 1: the soma signal fraction is estimated to $f_s$ ≈ 15% and the soma radius to $R_s$ ≈ 12 μm. Compared to SMEX, the extracellular parameters and $\tau_n$ are affected only slightly, while the neurite signal fraction and diffusivity are notably decreased. The increased model complexity resulted in several other minima, but they had worse fitting quality for all sub-ROIs except one for which the reported minimum and an alternative minimum could not be distinguished (ΔBIC < 1). Unsurprisingly, the alternative minimum persisted with eSANDIX (see next paragraph) for which it could generally not be disregarded based on fitting quality. The alternative parameter estimates are marked with an asterisk in Table 1: notable differences are that the neurites constitute more than two thirds of the signal, reducing the soma and especially extracellular signal fractions accordingly. The extracellular water exhibits essentially free diffusion $D_e \approx 2.4$ µm²/ms, and $R_s$ is reduced considerably to ≈ 7 µm, effectively switching those compartments' roles in terms of fast/slow diffusion.

Extending further to eSANDIX reproduces the apparent power-law at large b-values (Fig. 9, right) and increases the fitting quality relative to SANDIX significantly (ΔBIC > 10). Parameter estimates are given in Table 1: soma estimates, $\tau_n$, and extracellular parameters are only slightly altered compared to SANDIX except for the extracellular signal fraction, which is estimated to $f_e \approx 40\%$. The total neurite signal fraction is estimated to ≈ 45% with ≈ 8% of that stemming from impermeable neurites. The neurite diffusivity is estimated to be substantially smaller than with SMEX, $D_n \approx 0.6$ µm²/ms. Again, several other minima were encountered but these generally had worse fitting quality across all sub-ROIs with isolated cases of similar fitting quality. The aforementioned alternative minimum is an exception; it was in fact the global minimum in half of the sub-ROIs. The alternative estimates are marked with an asterisk in Table 1 and are similar to the equivalent estimates for SANDIX. The most notable difference being a reduced neurite diffusivity from roughly 0.9 to 0.6 µm²/ms. Compared to the other minimum, the neurite fraction is increased considerably to ≈ 72% with ≈ 5% of that stemming from impermeable neurites.

We explored the hypothesis that the apparent population of impermeable neurites stems from myelinated axons, which are known to be present in GM but much less than in WM (Jespersen et al., 2010). To this end we compared the signal from amygdala (see Fig. 2), which has a low myelin content (Jespersen et al., 2010), with that from a cortical sub-ROI with the largest apparent contribution from impermeable neurites. The signals are shown in Fig. 10 with eSANDIX fits. It is apparent that the amygdala signal attenuates faster consistent with less impermeable neurites. Accordingly, eSANDIX yielded a very small signal fraction from impermeable sticks of 0.3% relative to 5% for the cortical sub-ROI and was not justified over SANDIX based on BIC.

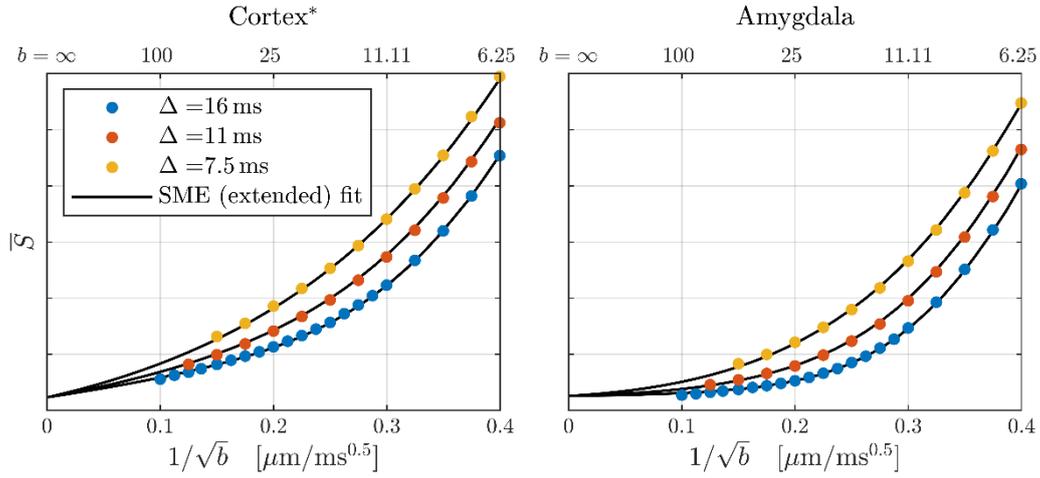

Fig. 10. Comparison of amygdala and a region of cortex (see Fig. 2 – third sub-ROI from left). The curves show fits of SMEX extended with both somas and impermeable neurites. Parameter estimates for the cortex region: $f_e$ = 38%, $D_e$ = 0.8 µm²/ms, $f_n$ = 38%, $f_n^{(imp)}$ = 5%, $D_n$ = 0.6 µm²/ms, $\tau_n$ = 4.5 ms, $f_s$ = 20%, $R_s$ = 13.6 µm, $f_{im}$ = 0.5%. For amygdala: $f_e$ = 44%, $D_e$ = 0.8 µm²/ms, $f_n$ = 48%, $f_n^{(imp)}$ = 0.3%, $D_n$ = 0.6 µm²/ms, $\tau_n$ = 3.5 ms, $f_s$ = 7%, $R_s$ = 11.7 µm, $f_{im}$ = 0.5%.

## 4. Discussion

We have investigated the lack of a stick power-law in GM (McKinnon et al., 2017; Veraart et al., 2020) using experiments on *ex vivo* rat brains and diffusion simulations in realistic neurons. The study has been centered on three previously suggested explanations for the power-law's absence (McKinnon et al., 2017; Özarslan et al., 2018; Veraart et al., 2019; Palombo et al., 2020): it is broken by (i) non-negligible neurite permeability or (ii) non-linear neurite morphology. Alternatively, the neurites exhibit the power-law but (iii) somas obscure the neurite signal.

Our strategy has relied on the identification of an experimental regime where neurites dominate the signal. In this regime, the hypotheses differ fundamentally in terms of power-law scaling and diffusion time dependence. This facilitated a qualitative discrimination of the hypotheses based on their experimental predictions. A quantitative evaluation of the corresponding microstructural models was enabled by incorporating exchange in the theory for powder-averaged dMRI experiments (Edén, 2003; Jespersen et al., 2013; Lasič et al., 2014), which has gained increasing attention in the recent years (Lampinen et al., 2017; McKinnon et al., 2017; Topgaard, 2017; Henriques et al., 2019; Afzali et al., 2020; Palombo et al., 2020). The predictions and models were applied to data obtained with an MR system capable of providing high-quality data at the appropriate diffusion weightings and gradient pulse times.

The main results are summarized here and discussed in the following sections. According to the simulations, impermeable neurites are well approximated by sticks. It has been shown that for Gaussian compartmental systems, the signal decreases with diffusion time at fixed b-value if exchange dominates the time dependence, and increases if structure dominates. This signature points to dominant exchange in the observed GM signal suggesting that the would-be GM power-law is broken by exchange. Interestingly, we do observe an apparent power-law in GM, but this is consistent with a small sub-population of impermeable neurites – plausibly myelinated axons – producing said power-law, while it is broken by exchange for the majority of neurites. The simulations and experimental data support that somas obscure the neurite behavior under typical experimental conditions, and as such both somas and exchange contributes to the lack of a GM power-law in practice.

The novelty of this study lies in isolating the neurite signal in GM, because so far the GM power-law's absence has not been attributable to exchange, somas, or morphology separately (McKinnon et al., 2017; Özarslan et al., 2018; Veraart et al., 2019, 2020). Evidence for exchange in GM has previously been reported (Pfeuffer et al., 1998, 1999; Quirk et al., 2003; Bai et al., 2018; Yang et al., 2018; Williamson et al., 2019; Jelescu and Novikov, 2020) but without consensus on intracellular residence times (Jelescu et al., 2020; Novikov, 2021). This study specifically points to fast neurite exchange. The findings also add to the discussion of exchange and structure as competing contributors to the MR signal's time dependence (Jelescu and Novikov, 2020; Lee et al., 2020b; Jelescu et al., 2021). In particular, the proposed exchange signature supports exchange to be the dominant effect in rat brain GM as also suggested by (Jelescu and Novikov, 2020; Jelescu et al., 2021).

### 4.1 The stick power-law withstands realistic neurite morphology

As an alternative to exchange (Özarslan et al., 2018) suggested neurite curvature and finite length cause the lack of a stick power-law in GM. Other deviations from sticks such as branching will also contribute. Using diffusion simulations in microscopy reconstructed neurons, we investigated whether these features are sufficiently pronounced in practice to break the power-law, and found that this is not the case: the neurites are approximated well by sticks for a broad range of gradient pulse times – ideally short δ and long Δ consistent with the theoretical predictions of (Özarslan et al., 2018) for small neurite curvature ($\delta D_0 \ll R_c$) though other aspects of the neuron morphology possibly contribute here. However, the neurite behavior is typically obscured by somas (and extracellular water) in the observable signal as proposed by (Palombo et al., 2020). These findings were independently reported by (Fang et al., 2020) also using simulations of microscopy reconstructed neurons simultaneously with the numerical part of this investigation (Olesen and Jespersen, 2020). The results are highly relevant because it validates modelling neurites as sticks in GM even if the stick power-law could not be observed in practice.

The simulations demonstrate that a strategy with large diffusion weighting similar to that applied to observe the stick power-law in WM (McKinnon et al., 2017; Veraart et al., 2019, 2020) can be applied in GM. However, even larger weighting might be required since the soma attenuation potentially is slower than that of extracellular water. Here, this is mitigated by using optimal gradient pulse times: $\delta$ and $\Delta$ are constrained by constant $\delta^2(\Delta-\delta/3)$ due to the available gradient strength, while short $\delta$, $\Delta$ maximize the soma attenuation, and sticks better approximate the neurites at short $\delta$, long $\Delta$. However, the substantial exchange indicated by the data breaks the would-be power-law for the (unmyelinated) neurites. Considering also that typical gradient strengths are insufficient to reproduce the experiment, an analysis of the optimum beyond what has been presented is unnecessary. We note, however, that the values will particularly depend on the gradient strength and the soma sizes in the sample (soma radii were 6-9 μm in the simulations).

## 4.2 Exchange dominates the time dependence

We have shown that if exchange dominates the time dependence, the signal decreases with increasing diffusion time at fixed b-value, while it increases if structure dominates. The observed decreasing signal thus supports the initial suggestion by (McKinnon et al., 2017) that exchange breaks the GM power-law. Nevertheless, we do see an approximate stick power-law at large $\Delta$, but this is consistent with a small population of myelinated axons. This is corroborated by the significantly less pronounced power-law in amygdala, which has a lower myelin content. Accordingly, this ROI yields an apparent signal fraction from impermeable sticks of less than 1% compared to cortical values up to ≈ 5%. Overall the results thus support that neurite exchange is non-negligible and breaks the stick power-law for the majority of neurites in GM but any myelinated neurites will unsurprisingly still exhibit a power-law.

The data is also consistent with a distinct soma compartment. Unaccounted for, somas will thus likely bias data interpretation in typical experiments as proposed by (Palombo et al., 2020); for instance, somas probably contribute with exchange to the absence of a stick power-law reported by (McKinnon et al., 2017; Veraart et al., 2020). As such, with significant exchange, there is no (very little) power-law for the somas to obscure in the first place, but they will obscure the neurite behavior.

The time dependence signature is derived from models in the GPA. As such, while time dependent diffusion cannot explain our data, we can in principle not exclude that non-exchanging non-Gaussian compartments can. However, we consider exchange the most plausible explanation because of the weak time dependence of the GM diffusivity relative to kurtosis, as reported also by (Pyatigorskaya et al., 2014; Aggarwal et al., 2020; Jelescu et al., 2021). This aligns with the Kärger model's predictions with the diffusivity's minor time dependence possibly attributable to somas as incorporated in the SMEX extensions. We also note that taking higher-order effects into account for the somas using the MCF approach (Callaghan, 1997; Grebenkov, 2008) did not affect the results eliminating the concern for that compartment. This is also supported by the simulations, which intrinsically include higher-order effects, for the somas and also the neurites.

Related to this discussion, (Henriques et al., 2020) found appreciable intracompartmental kurtosis throughout the rat brain. While this apparently supports structure as a contributor, it should be noted that the employed Correlation Tensor Imaging (CTI), an otherwise model-free framework, assumes negligible exchange. The results are therefore not inconsistent with a combination of the effects; exchange could be a contributing factor to microscopic kurtosis measured via CTI. As such, extending CTI to include exchange effects might further increase the relevance of the framework for GM modelling.

While exchange appears to dominate the time dependence in rat brain at the time scale probed here, structural effects should be expected to dominate at sufficiently short times. An initial increase in kurtosis followed by a decrease as observed by (Pyatigorskaya et al., 2014; Aggarwal et al., 2020) in rodent brain supports this notion. Similar behavior was observed in porcine spinal cord by (Jespersen et al., 2018) with the turning point at longer diffusion time. The relative contributions from exchange and structure could vary between species and notably might vary from rat to human brain. Accordingly, (Lee et al., 2020b) analyzed D(t) and K(t) in human GM and suggested structure as the dominant contributor, although speculating that the data was in a "cross-over-regime" between the two effects. Interestingly, they also report a signal decrease with increasing diffusion time at fixed b-value – our proposed signature for exchange. The decrease is more subtle than reported here though, as expected due to the use of much smaller b-values (see section 2.3).

### 4.3 Model parameter estimates

Modelling both exchange and somas is an ambitious program from a quantification viewpoint because of the many parameters and resulting multiple minima. Nevertheless, we have attempted to quantify the corresponding models with key takeaway that neurites are in fast exchange with extracellular water (mean neurite residence time $\tau_n \approx 4$ ms). Such fast exchange could be problematic from a modelling viewpoint because the diffusion must be barrier-limited for the Kärger model to apply (Fieremans et al., 2010), but the necessary time scales for this to be achieved remains to be established. Furthermore, a relevant modelling alternative – outside explicit Monte Carlo simulations – has yet to be proposed (Novikov, 2021).

There is no consensus regarding intra-cellular residence time in GM; estimates span values from roughly 10 ms to considerably longer than 100 ms as recently reviewed by (Jelescu et al., 2020; Novikov, 2021). The large span of residence times could indicate the presence of both fast and slowly exchanging components as proposed by (Nilsson et al., 2013). They speculated that astrocytes could be the fast component with neurons being slowly exchanging. On the other hand, (Yang et al., 2018) later found astrocytes and neurons to have similar residence times on the order of 500 ms using cells grown on microbeads. Our findings suggest that neurons are not a slowly exchanging component because then a stick power-law could be expected in all GM and not only regions with myelin. In comparison to the literature, our estimate of roughly 4 ms is in the lowest end with the closest estimate being 9±2 ms by (Williamson et al., 2019) (at temperature 25 °C). Their study and ours employed fixed tissue which potentially contributes to explaining the fast exchange because fixation related protein cross-linking could increase cell membrane permeabilities (Shepherd et al., 2009). The neurite permeability can be estimated using its relation to exchange rate and surface to volume ratio (Ash et al., 1978) in the well-mixed regime (Jespersen et al., 2005): for a typical neurite radius of 1 μm (Palombo et al., 2021) with residence time 4 ms, this results in a permeability of 125 μm/s. This is similar to that of red blood cells, which are known to be highly permeable – in the range 49-112 μm/s at 37 °C (Benga et al., 2009).

The signal fractions reflect, apart from volume, also unknown T2 and T1 weighting if these differ amongst the compartments. The estimates should therefore only be taken as an indication of volume when compared with histology. Nevertheless, it is encouraging that the SM with exchange yields a large neurite signal fraction up to 78% (dependent on the specific model variant) and also reasonable extracellular and soma signal fractions. This is in the context that the neurite contribution typically has been estimated to be much smaller in GM than in WM using models without exchange (Jespersen et al., 2010; Lampinen et al., 2019, 2020; Palombo et al., 2020; Afzali et al., 2021) despite its volume fraction being quite high. We reiterate from (Jelescu et al., 2020) that histology suggests a neurite volume fraction of 40-75% with extracellular water comprising 15-30% and somas 10-40% (Bondareff and Pysh, 1968; Spocter et al., 2012; Motta et al., 2019). Regarding the two minima of the extended exchange models, the signal fractions do present a case for the alternative minimum because the extracellular signal fraction is generally lower (≈ 15% relative to ≈ 35%) and the neurite signal fraction higher (≈ 75% relative to ≈ 50%). On the other hand, the alternative minimum suggests a soma signal fraction of ≈ 10% relative to ≈ 15% which is arguably small.

The compartmental diffusivities are unknown, so the minima's differing extracellular diffusivities do not make a case for either of them. The corresponding difference in the soma radius is discussed in section 4.4. The neurite diffusivity is estimated to be smaller than extra-neurite diffusivities in all cases except the basic SMEX model. This is slightly surprising in the context of the recent branch-debate in WM modelling (closely related to the relative values of the intra-axonal and extracellular diffusivities), because the consensus is that the intra-axonal diffusivity is largest (Fieremans et al., 2011; Dhital et al., 2018, 2019; Jespersen et al., 2018; Kunz et al., 2018; McKinnon et al., 2018; Lampinen et al., 2020; Olesen et al., 2021). However, this could be different in GM and may reflect the different structure of GM neurites relative to myelinated axons.

In the context of previous reports, especially *ex vivo*, of a significant immobile water fraction, (Dhital et al., 2018; Veraart et al., 2019; Tax et al., 2020) recently reported upper limits on the order of 1% in human brain *in vivo* using isotropic encoding and/or very high diffusion weighting. We found an apparent immobile water fraction below 1% in GM with eSANDIX, which accounted well for the data at large b-values. This is consistent with the findings of (Tax et al., 2020) who reported a <2% contribution from immobile water in the human cerebellar cortex. In WM, they reported an upper limit of less than 1% consistent with our findings in corpus callosum. These results suggest that immobile water can be neglected in these regions even for *ex vivo* experiments without large diffusion weighting.

## 4.4 The soma MR radius

The soma MR radius overestimates the true radius because of volume weighting (in proportion to $R^3$) and since it is probed indirectly through the diffusivity, which is not linearly dependent on R. These are well-known effects in the context of axon radii, which are very small ($\delta D \gg R^2$). In that limit, the diffusivity is proportional to $R^4$ for $\delta = \Delta$, or proportional to $R^2$ for $\delta \ll \Delta$. Combined with a factor $R^2$ from volume weighting of cylinders, yields the well-known formulas for the axon MR radius: $(\langle R^6 \rangle / \langle R^2 \rangle)^{1/4}$ or $(\langle R^4 \rangle / \langle R^2 \rangle)^{1/2}$ respectively (Burcaw et al., 2015). Similar formulas apply for small somas: $(\langle R^7 \rangle / \langle R^3 \rangle)^{1/4}$ or $(\langle R^5 \rangle / \langle R^3 \rangle)^{1/2}$. However, somas can typically not be considered small, the effect of which is apparent in the narrow-pulse limit $\delta D \ll R^2$, where the soma diffusivity is (Balinov et al., 1993)

$$D_s(\Delta, R, D_0) \approx \frac{2R^2}{\Delta} \sum_{m=1} \frac{\alpha_m^{-2}}{\alpha_m^2 - 2}\left(1 - \exp\left(-\frac{\alpha_m^2 D_0 \Delta}{R^2}\right)\right). \quad (11)$$

Here, $\alpha_m$ are the roots of $J_{3/2}(\alpha) = \alpha J_{5/2}(\alpha)$ and $J_n$ are Bessel functions of the first kind. The diffusivity is proportional to $R^2$ for small radii but the exponential factor increasingly curves this for larger radii towards the asymptote $D_0$. Using the general sphere diffusivity, and a gamma distribution for the soma radius with mean 7.1 µm and standard deviation 3.6 µm matching those of rat pyramidal cells from NeuroMorpho.org (Palombo et al., 2021) gives a rough estimate of the expected MR radius of 11.1 µm – a considerable overestimation. Note that this value deviates substantially from the approximation $(\langle R^5 \rangle / \langle R^3 \rangle)^{1/2} \approx 13.5$ µm for small somas.

The MR radius is also affected by being estimated simultaneously with other signal components such as extracellular water; somas with sizes resulting in diffusivities similar to the extracellular diffusivity are difficult to distinguish from that component unless diffusion times are sufficiently varied. Depending on the diffusivity assigned to the extracellular compartment, the MR radius can correspondingly over- or underestimate the true MR radius. This effect is small compared to the effects discussed above but is notable for our employed acquisition: using eSANDIX to simulate the signal with parameters from Table 1 and replacing the soma radius with the aforementioned distribution, the radius should ideally be estimated to 11.1 µm, but is overestimated to 11.8 µm.

In summary, the soma size can be expected to be considerably overestimated. Regarding the competing minima, this points to the one with larger $R_s$. Using the simulated signal from before and introducing noise results in a competitive minimum with smaller $R_s$ consistent with the reported fitting behavior. It is plausible that this degeneracy can be lifted with dedicated experiments employing for instance double wave vector (Cory et al., 1990; Shemesh et al., 2016) or b-tensor encoding (Westin et al., 2016; Lampinen et al., 2020), as has been the case for the reminiscent branch issue in WM modelling (Coelho et al., 2019; Reisert et al., 2019; Olesen et al., 2021). This has recently been explored in the context of SANDI (Afzali et al., 2021; Gyori et al., 2021) but without incorporating exchange.

### 4.5 Clinical implications

Brain development, learning, and aging are associated with dynamic changes in cortical microstructure on different time scales, as are major neurodevelopmental and neurodegenerative disorders. These microstructural changes parallel the formation, maintenance, and loss of key components of the cortical, computational machinery: the synapses, dendrites, soma, and axons of cortical neurons. The ability to non-invasively separate and quantify these components over time is therefore crucial for our efforts to identify the etiology and disease mechanisms underlying brain disorders, to diagnose them earlier, and to detect changes in their progression during drug development. In view of this promise, we hope our incremental steps toward disentangling the origin of cortical diffusion signals may ultimately benefit patients with severe brain disorders.

Our findings could be utilized as a potential new contrast for exchange between (ad hoc) neurites and extracellular space. Although beyond the scope of this work, it seems plausible that voxel wise mapping of exchange rates would reveal features proportional to the permeability of the neurites. This could impact the investigation of neural injury, e.g., stroke, where neurite beading effects are related to ionic imbalances and ensuing transport disruptions. Furthermore, such metrics could be useful for other types of disease where exchange rates may differ, e.g. cancer.

With respect to the clinical implementation of SMEX and derived models, it should be noted that it requires at minimum a second "SANDI"-like measurement with another diffusion time. It could again be argued that this is achievable in principle. Alternatively, an experimental design with many diffusion times rather than gradient values could benefit the approximation of the neurite exchange rate (Meier et al., 2003). A potentially more feasible alternative to applying a full model is accommodated by Eq. (7). It follows that the signal at a later time $t$ relative to $t_0$ is given by

$$\log \frac{\overline{S}(t_0 + t)}{\overline{S}(t_0)} \sim -r_n \left(1 - \frac{2}{bD_e} + 2t_0 \frac{r_e}{bD_e}\right) t - r_n \frac{r_e}{bD_e} t^2$$
$$\approx -r_n \left(1 + 2t_0 \frac{r_e}{bD_e}\right) t - r_n \frac{r_e}{bD_e} t^2 \,. \tag{12}$$

That is, the neurite exchange rate can be obtained from linearly combining the coefficients of the second-order polynomial, which approximates the signal's time-dependence at fixed b-value for small time differences. Part of this insight was presented by (Meier et al., 2003) in the limit of slow exchange for which the right-hand-side reduces to the neurite exchange rate. Eq. (13) is an increasingly good approximation for large b-values and slow exchange/short time ($r_e t/bD_e \ll 1$). The required measurements are reduced to one b-value with multiple diffusion times, but the expression relies on the narrow-pulse approximation and a sufficiently large b-value, which are challenging to fulfill simultaneously. The narrow-pulse requirement can be alleviated by instead numerically solving the generalized rate equations (Ning et al., 2018). This lacks the benefit of an analytical expression like Eq. (13), but has the same key feature of low sensitivity towards other parameters than $r_n$ at large b-values.

## 5. Conclusion

We used experiments on *ex vivo* rat brains and diffusion simulations in microscopy reconstructed neurons to explore the functional characteristics of the GM dMRI signal. The simulations demonstrate that neurite morphology does not break the stick power-law, but that this typically is obscured by somas. Even so, the neurites can be made to dominate the signal with carefully chosen gradient pulse times and strong gradients, facilitating observation of a potential stick power-law. Analytically analyzing Gaussian compartmental systems, we identified the sign of the MR signal's time dependence at fixed b-value as a functional signature for whether exchange or structure dominates the time dependence. Applying these results to data from *ex vivo* rat brain acquired with a state-of-the-art system, we found evidence for substantial exchange but also an apparent stick power-law in GM. These observations are consistent with non-negligible neurite exchange plus a small population of myelinated, impermeable axons. A quantitative description of the data was facilitated by incorporating exchange in the SM and SANDI models. The data support including both neurite exchange and a soma compartment in a potential GM model.

## Acknowledgements

We thank Beatriz Cardoso and Cristina Chavarrías for preparing the specimens and assistance with the experiments. The graphical representation of the models in Fig. 1 was created with BioRender.com. SJ, LØ, and JO are supported by the Danish National Research Foundation (CFIN), and the Danish Ministry of Science, Innovation, and Education (MINDLab). Additionally, LØ and JO are supported by the VELUX Foundation (ARCADIA, grant no. 00015963). NS was supported in part by the European Research Council (ERC) (agreement No. 679058). The authors acknowledge the vivarium of the Champalimaud Centre for the Unknown, a facility of CONGENTO which is a research infrastructure co-financed by Lisboa Regional Operational Programme (Lisboa 2020), under the PORTUGAL 2020 Partnership Agreement through the European Regional Development Fund (ERDF) and Fundação para a Ciência e Tecnologia (Portugal), project LISBOA-01-0145-FEDER-022170.

## Appendix A

Here, we present supporting data from two preliminary experiments (specimen 2 and 3) and one repetition of the experiment described in the main text (specimen 4). The supporting data exhibits the same features as the main dataset thereby demonstrating reproducibility.

The preliminary experiments were carried out as detailed in section 2.5 with the specific b-values and pulse separations differing. For specimen 2, data was recorded only for Δ = 16 ms with b-values identical to those of the main experiment except that the b-value at 0.1 ms/µm² was omitted. For specimen 3, pulse separations Δ = 16 and 8 ms were used with the following shared b-values: b = 0.5, 1, 2, 3, 4, and 5 ms/µm² and larger b-values in terms of $b^{-1/2}$: 0.375, 0.325, 0.3, 0.275, 0.25, 0.225, 0.2, 0.175, and 0.15 µm/ms$^{1/2}$. For Δ = 16 ms, the following additional b-values in terms of $b^{-1/2}$ were used: 0.1875, 0.1625, 0.1375, 0.125, 0.1125, and 0.1 µm/ms$^{1/2}$.

The GM signals as represented by cortical ROIs are shown in Fig. A.1. The cortical ROIs are placed as in Fig. 2. The data are qualitatively in agreement with the main dataset sharing the follow signature features: an apparent stick power-law at very large b-values and a clearly decreasing signal as a function of time.

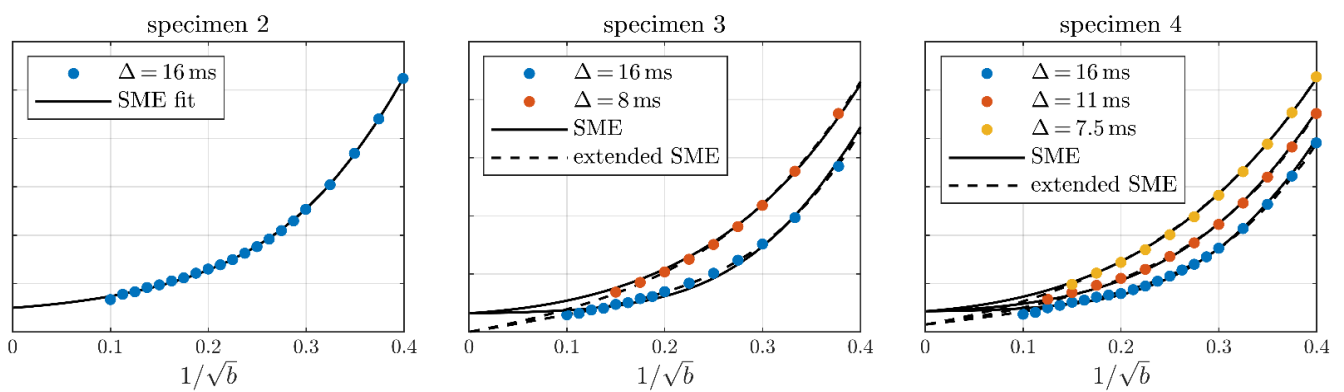

Fig. A.1. The large-b GM signal from three supporting experiments on other brains. Data is shown with markers and fits of the SMEX and the extended SMEX (including somas and non-exchanging neurites) models are shown with solid and dashed curves respectively. In the case of specimen 2, only SMEX was fitted to the data.

Parameters for fits of SMEX and eSANDIX are given in table A.1. SMEX accounts for the time dependence but not the power-law, which is accounted for by eSANDIX. The estimates indicate quantitative reproducibility as they are consistent with those of the main dataset even though estimates for the preliminary datasets deviates notably likely due to the reduced data. For specimen 2, the estimates differ but a fair comparison would be with the parameters provided in the caption of Fig. 8 where SMEX is fitted only to the subset of data with Δ = 16 ms – these are in agreement.

|  | $f_e$ | $D_e$ | $f_n$ | $f_n^{(imp)}$ | $D_n$ | $\tau_n$ | $f_s$ | $R_s$ | $f_{im}$ |
|---|---|---|---|---|---|---|---|---|---|
| Spec. 2, SMEX | 50-65 | 0.5-0.6 | 47-58 | - | 1.1-1.8 | 5.6-10.7 | - | - | 0.7-1.2 |
| Spec. 3, SMEX | 34-44 | 0.7-1.1 | 55-66 | - | 1.0-2.0 | 2.7-4.5 | - | - | 0.5-0.8 |
| Spec. 4, SMEX | 27-41 | 0.8-1.1 | 58-71 | - | 0.8-1.6 | 4.2-5.7 | - | - | 0.9-1.3 |
| Spec. 4, SANDIX | 26-40 | 0.8-1.0 | 49-59 | - | 0.5-1.1 | 4.1-5.8 | 10-16 | 9.2-13.6 | 0.8-1.3 |
| Spec. 4, eSANDIX | 34-50 | 0.8-0.9 | 32-46 | 3-7 | 0.3-0.6 | 3.9-4.8 | 11-16 | 9.6-13.3 | 0.1-0.7 |
| Spec. 4, SANDIX* | 11-26 | 1.3-2.5 | 67-75 | - | 0.6-1.0 | 3.4-4.6 | 7-13 | 7.5-8.5 | 1.0-1.5 |
| Spec. 4, eSANDIX* | 15-41 | 1.2-2.3 | 48-67 | 3-8 | 0.4-0.8 | 3.5-4.2 | 6-13 | 5.7-7.1 | 0.3-0.6 |

Table A.1. Model parameter estimates for exchange models from fitting to the GM signals from the supporting data. Diffusivities are given in µm²/ms, $\tau_n$ in ms, and $R_s$ in µm.